\documentclass{article}
\usepackage{amsmath}
\usepackage{amssymb}
\usepackage{graphicx}
\usepackage{wrapfig}
\usepackage[percent]{overpic}
\newcommand{\D}{\,\mathrm{d}}
\newcommand{\I}{\mathrm{i}}
\newcommand{\pd}[2]{\ensuremath{\frac{\partial #1}{\partial #2}}}
\newcommand{\pdl}[2]{\ensuremath{\partial #1 / \partial #2}}
\newcommand{\pdc}[3]{\ensuremath{\left(\frac{\partial #1}{\partial #2}\right)_{#3}}}
\newcommand{\etc}{\textit{etc.}}
\newcommand{\cf}{\textit{cf.}}

\newcommand{\vnb}{\ensuremath{\mathbf{v}_\text{n}}}

\newcommand{\ept}{\ensuremath{\widetilde{\varepsilon}}}
\newcommand{\et}{\ensuremath{\widetilde{\mathcal{E}}}}
\newcommand{\kb}{\ensuremath{\mathbf{k}}}

\newcommand{\vst}{\ensuremath{v_\text{s}'}}
\newcommand{\vsb}{\ensuremath{\mathbf{v}_\text{s}}}
\newcommand{\vsbt}{\ensuremath{\mathbf{v}}_\text{s}'}
\newcommand{\vsbd}{\ensuremath{\dot{\mathbf{v}}_\text{s}}}

\newcommand{\wbt}{\ensuremath{\mathbf{w}'}}

\newcommand{\wb}{\ensuremath{\mathbf{w}}}
\newcommand{\pb}{\ensuremath{\mathbf{p}}}
\newcommand{\jb}{\ensuremath{\mathbf{j}}}
\newcommand{\jn}{\ensuremath{j_0}}
\newcommand{\jnb}{\ensuremath{\mathbf{j}_0}}

\newcommand{\jntx}{\ensuremath{j_{0x}'}}
\newcommand{\jnty}{\ensuremath{j_{0y}'}}

\newcommand{\jnbt}{\ensuremath{\mathbf{j}_0'}}

\newcommand{\mut}{\ensuremath{\mu'}}
\newcommand{\rt}{\ensuremath{\rho'}}

\newcommand{\pdvlr}{\ensuremath{\frac{\partial v_L}{\partial \rho}}}

\author{A.F.\,Andreev\footnote{Deceased March 15, 2023.}$\ ^a$
, L.A.\,Melnikovsky$\,^{b,a}$
\\
$^a\,$P.L.\,Kapitza Institute for Physical Problems, Moscow, Russia\\
$^b\,$Weizmann Institute of Science, Rehovot, Israel
}
\title{Overcritical Fermi Superfluids}
\date{}
\begin{document}
\maketitle

\begin{abstract}
Superfluidity in Fermi systems is not destroyed by a flow exceeding the Landau velocity threshold. The overcritical state acquires normal component even at zero temperature. We explore peculiar hydrodynamics of this system and discover two sound modes, for which the explicit universal dispersion relations are provided.
\end{abstract}

\section{Introduction}
Superfluid hydrodynamics is notable for its two velocities and two ``components'' with temperature dependent densities. In pure superfluids (pristine $^4$He and $^3$He) the normal component density vanishes at zero temperature. Quasiparticles of $^3$He in superfluid $^4$He\,--$^3$He solutions contribute to the normal density and keep it finite even at absolute zero.

According to Landau criterion, the velocity of superflow with respect to the reservoir walls is limited by a critical value $v_L$. Beyond this limit the quasiparticles with negative energy are spontaneously created. If some mechanism puts a bound on the quasiparticle population, then a phase transition occurs at the Landau critical velocity. The nature of the emerging overcritical state depends on the quasiparticle statistics. In the $^4$He Bose case, in presence of suitable repulsive interaction, the fluid may become nonuniform \cite{pit,self}. Overcritical Fermi superfluid is stabilized by the Pauli principle. Equilibrium at zero temperature corresponds to a non-vanishing normal component (like in $^4$He\,--$^3$He solutions) formed by fully occupied negative energy Bogolyubov quasiparticle levels \cite{volovik}. Indeed, the superfluidity in Fermionic system seems to survive the passing of Landau critical velocity \cite{lancaster1,lancaster2}.

Below we explore the exact nonlinear superfluid hydrodynamics \cite{khalat} and particularly investigate the sound propagation in overcritical $^3$He-B at zero temperature for small over-speed (and therefore for dilute normal component). Like the usual $^4$He superfluid at rest, this system has two acoustic modes:
\begin{enumerate}
    \item The fast one is effectively the conventional sound, slightly entrained downstream by the quasiparticle flow.
    \item The slow mode, naturally referred to as the second sound, strongly depends on the overspeed and is essentially anisotropic. It is ``supported'' by the normal component and its velocity is close to the velocity of the latter. Just like the conventional second sound, the fluid density remains constant in these waves, but {\em{unlike}} the conventional second sound, no temperature oscillations are associated with them.
\end{enumerate}


\section{Thermodynamics of Bogolyubov-Fermi Gas}
For simplicity, we consider an isotropic gap Fermi superfluid at zero temperature. A degenerate kind of thermodynamics is used in this case. The energy of Bogolyubov quasiparticles in $^3$He-B is given by the usual dispersion law
\begin{equation*}
\varepsilon(\pb)=\sqrt{(p-p_F)^2 v_F^2 + \Delta^2}\approx\Delta+\frac{(p-p_F)^2}{2m},
\end{equation*}
where $p_F$ is the Fermi momentum, $v_F$ is the Fermi velocity, $\Delta$ is the superfluid energy gap, and the effective mass is defined as $m=\Delta/v_F^2$. It is shown below \eqref{jofu} that equilibrium number of the quasiparticles is proportional to a high power of the over-speed. This justifies the disregard of quasiparticle interaction effects in the main approximation (particularly the Fermi-liquid and the gap self-consistency corrections, recursive Cooper pairing of Bogolyubov quasiparticles \etc) and allows us to assume that the dispersion parameters $\Delta$ and $m$ depend on the fluid density $\rho$ alone.

\begin{figure}[h]
\begin{center}
\includegraphics[width=0.8\linewidth]{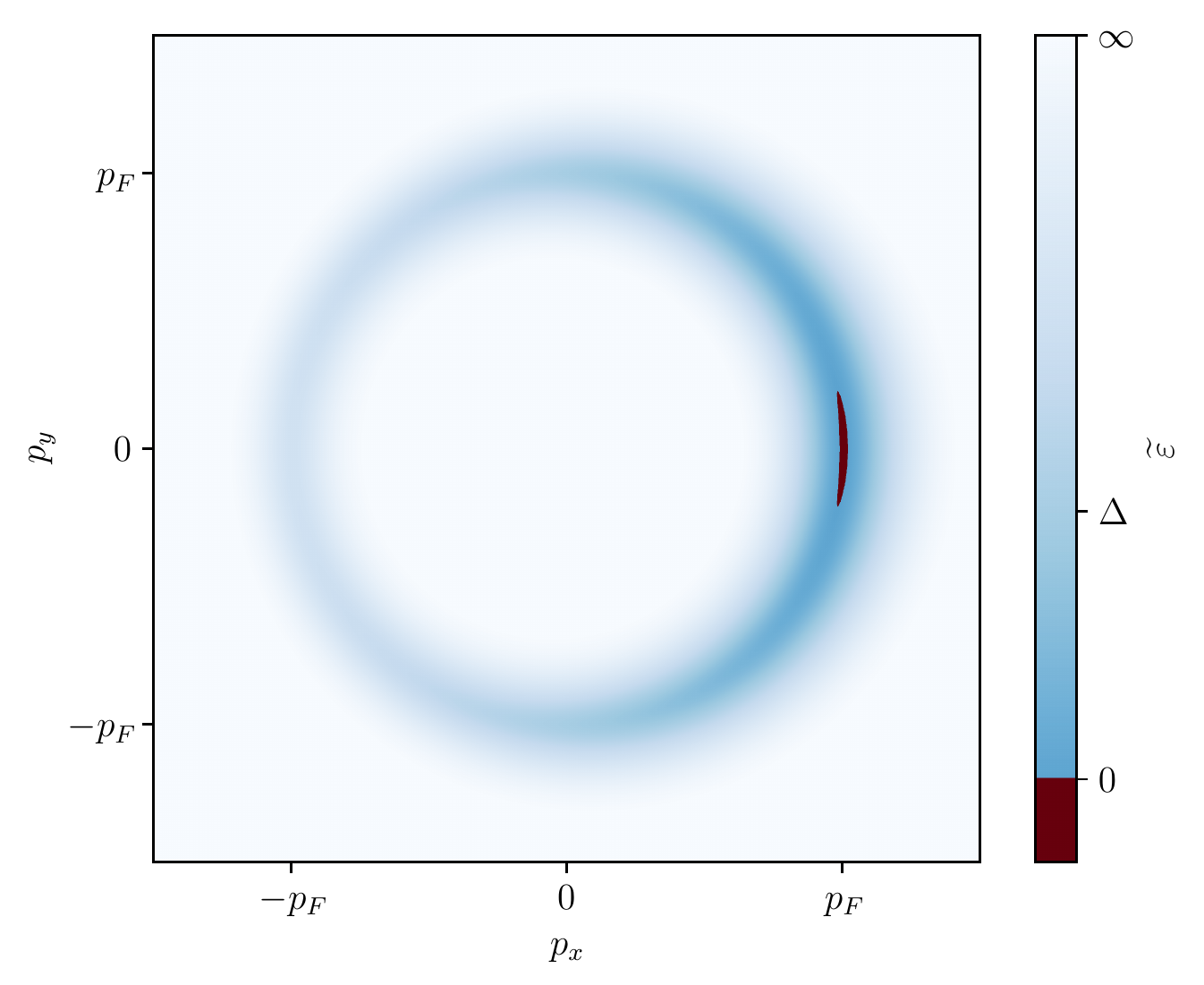}
\end{center}
\caption{Reduced energy \ept\ of the excitations \eqref{eq-ept} in overcritical Fermi superfluid. Bogolyubov-Fermi normal component is formed by the negative energy states.}
\label{ept}
\end{figure}
Introduce the reduced quasiparticle energy $\ept(\pb)\equiv  \varepsilon(\pb)-(\pb\wb)$, where $\wb=\vnb-\vsb$ is the relative velocity of the normal and superfluid motion. The inequality $\ept(\pb) < 0 $ has solutions if the value of the relative velocity is above the Landau threshold $w>v_L$. The latter is much smaller (see Section \ref{incompress} below) than the Fermi velocity $v_L \ll v_F$ and is given by the expression
\begin{equation*}
v_L = \frac{p_F}{m} \left(\sqrt{1+\frac{2m\Delta}{p_F^2}}-1\right)\approx
\frac{\Delta}{p_F}-\frac{m\Delta^2}{2p_F^3}.
\end{equation*}
If we further assume the over-critical regime $w>v_L$ but small over-speed $w-v_L\equiv u \ll v_L$, then the reduced energy $\ept(\pb)$ can be expanded (similar transformation can be applied to the rotons \cite{BEC}) near its minimum $p_0=p_F+mw$ as follows:
\begin{equation}
\label{eq-ept}
\ept=-u p_F
+\frac{f_\parallel^2}{2m}+\frac{f_{\perp\vphantom{\parallel}}^2 w}{2p_F}=
-u p_F + \frac{g^2}{2m},
\end{equation}
where we denote
\begin{align*}
\mathbf{f}=\mathbf{p}-\wb\left(\frac{p_0}{w}+m\right)&,&
f_\parallel&=\frac{(\mathbf{f}\wb)}{w},&
f^2&=f_\parallel^2+f_{\perp\vphantom{\parallel}}^2,\\
g_{\perp\vphantom{\parallel}} = f_{\perp\vphantom{\parallel}} \sqrt{mv_L/p_F}&,&
g_\parallel &= f_\parallel,&
g^2&=g_\parallel^2+g_{\perp\vphantom{\parallel}}^2.
\end{align*}

At zero temperature all states with negative reduced energy \ept\ are occupied, see Fig.\ref{ept}, thus forming a filled ellipsoid inside the Bogolyubov-Fermi surface. This simple picture holds at low temperature $T \ll T_0$, where $T_0 = up_F$ is an effective Bogolyubov-Fermi energy. By integrating \ept\ over the populated states we get the Legendre transform of the total quasiparticle energy density:
\begin{equation*}
\et =
\int\limits_{\ept<0} \ept \frac{2 \D \mathbf{p}}{(2\pi\hbar)^3} =
\frac{2 p_F}{(2\pi\hbar)^3m v_L} \int\limits_{\ept<0} \ept \D \mathbf{g}=
-\frac{4 \sqrt{2 p_F^7 m} }{15 \pi^2 \hbar^3 v_L} u^{5/2}\equiv
-\gamma u^{5/2}.
\end{equation*}
The fluid energy density in the condensate frame of reference is
\begin{equation*}
E_0=E_0(\rho) + \et + \jnb\wb, \quad 
\D E_0 = \mu\D\rho + \wb\D\jnb,
\end{equation*}
where $\jnb=2\int \mathbf{p} \D \mathbf{p}/(2\pi\hbar)^3$ is the quasiparticle momentum density, $\rho$ is the mass density, and $\mu$ is the chemical potential. It immediately follows, that
\begin{gather}
\label{jofu}
\jn=-\pdc{\et}{w}{\rho}=-\pdc{\et}{u}{\rho}=\frac{5 \gamma}{2} u^{3/2},\quad
u=\left(\frac{2\jn}{5\gamma}\right)^{2/3},
\\
E_0=E_0(\rho)+ \jn v_L + \frac{3}{5} \left(\frac{2}{5\gamma}\right)^{2/3}\jn^{5/3},
\notag\\
\label{thermodynamics}
\wb=\frac{\jnb}{\jn}\left(v_L + \left(\frac{2 \jn}{5 \gamma}\right)^{2/3}\right)
,\quad
\mu=\mu_0 + \jn \pdvlr - \left(\frac{2\jn}{5\gamma}\right)^{5/3}\pd{\gamma}{\rho},
\end{gather}
where $\mu_0=\pdl{E_0(\rho)}{\rho}$ is the chemical potential of the fluid at rest. The overcritical state is thermodynamically stable, this is straightforward to verify Eqs.\eqref{jofu}-\eqref{thermodynamics} against the thermodynamic inequalities, obtained in \cite{superstab}.

\section{Hydrodynamics of Overcritical Fermi Superfluid}
To investigate the properties of the system in the overcritical region one must use exact equations of nonlinear Landau two-fluid hydrodynamics \cite{khalat}.
Without the thermal terms we have:
\begin{align}
\label{H1}
0 &= \dot{\rho} + (\nabla \jb), \\
\label{H2}
0 &= \vsbd + \nabla \left( \frac{\vsb^2}{2} + \mu \right),\\
\label{H3}
0 &= \dot{\jb} + \vsb (\nabla \jb) + (\jb \nabla) \vsb + \jnb (\nabla \vnb) + (\vnb \nabla) \jnb + \nabla p,
\end{align}
where $\jb=\jnb+\rho\vsb$ is the momentum density and
\begin{equation}
\label{pressure}
p=\mu\rho + \jnb\wb - E_0
\end{equation}
is the pressure. The equations \eqref{H1}-\eqref{H3} can be linearized around the static uniform overcritical state. This is done by decomposing all dynamic local variables ($\rho+\rt$ for the density, $\mu+\mut$ for the chemical potential, $\wb+\wbt$ for the relative velocity, \etc) as the sums of the corresponding equilibrium time-independent average values and oscillating terms denoted by the primed letters. These oscillating terms are assumed to be small, we retain only the first order terms in them. Their temporal and spatial dependencies in a plane wave are harmonic $\propto \exp(\I \kb \mathbf{r} - \I \omega t)$ with a wave vector \kb\ and a frequency $\omega$.

We use the ``average frame of reference of the superfluid component''. This means the superfluid velocity has just one term $0+\vsbt$, the average normal velocity $\vnb=\wb$ is slightly above the Landau critical velocity $v_L$, and the average mass flux is due to the quasiparticles only $\jnb=\jb$, $\jb'=\jnbt + \rho\vsbt$.

The primed variables are tied by the equations \eqref{thermodynamics} and \eqref{pressure}:
\begin{equation}
\label{primes}
\begin{aligned}
\wbt&=
\frac{w}{j} \jnbt
-\frac{w + 2 v_L}{3 j}\frac{\jb(\jb\jnbt)}{j^2}
+ \frac{\jb}{j} A \rt,\\
\mut&=
A\frac{(\jb\jnbt)}{j}
+ \left(\frac{c^2}{\rho} + B \right)\rt,\\
p'&=\rho \mu' + (\jb \wbt),
\end{aligned}\end{equation}
where
\begin{align*}
A &= \pdvlr  - \frac{2 u}{3\gamma}\pd{\gamma}{\rho}, \qquad c^2=\rho\pd{^2E_0(\rho)}{\rho^2}, \\
B &= j \pd{^2 v_L}{\rho^2} - u^{5/2}\pd{^2\gamma}{\rho^2} + \frac{5 u^{5/2}}{3\gamma}\left(\pd{\gamma}{\rho}\right)^2.
\end{align*}
Here $c$ is the sound velocity in the unperturbed liquid.

From Eq.\eqref{H2} it follows that the superfluid velocity is collinear with the wave vector $\vsbt \parallel \kb$. Entire problem is therefore restricted to the span of $\{\kb,\jb\}$ and the flow is effectively two-dimensional. Separating the components of the Eq.\eqref{H3} along \kb\ and along \jb\ and substituting the Eqs.\eqref{primes} into the Eqs.\eqref{H1}-\eqref{H3} we obtain
\begin{equation}
\label{Matrix}
\begin{aligned}
\omega \rt &= && &k\rho&\vst && & +k&\frac{(\kb \jnbt)}{k}, \\
\omega \vst &= & k\left(\frac{c^2}{\rho} + B\right)&\rt &&& + kA&\frac{(\jb\jnbt)}{j},&&\\
\omega \frac{(\jb\jnbt)}{j} &=
&2 A jk \cos\theta&\rt
&+  C jk&\vst 
&+ Dk &\frac{(\jb\jnbt)}{j}
&+ wk&\frac{(\kb\jnbt)}{k},\\
\omega \frac{(\kb \jnbt)}{k} &=
&A C j k  &\rt
&+ 2 jk \cos\theta &\vst 
&+ Fk&\frac{(\jb\jnbt)}{j}
&+ 2 wk \cos\theta &\frac{(\kb \jnbt)}{k},
\end{aligned}
\end{equation}
where we denote
\begin{align*}
\cos\theta &=\frac{(\jb\kb)}{jk},
& C &= \cos^2\theta+1,\\
D &= \frac{4u}{3}\cos\theta,
& F &= \frac{2u}{3}C-w\cos^2\theta.
\end{align*}

The characteristic equation of \eqref{Matrix} is of the fourth order. Approximate eigendecomposition will be done separately for ``fast'' and ``slow'' modes.

\subsection{First Sound $\omega \sim ck$}
The usual sound dispersion relation in the fluid at rest is $\omega=\pm ck$. The influence of the normal component can be taken into account perturbatively from the power series expansion of the characteristic polynomial for the system \eqref{Matrix} around $\pm ck$. In the first order we get
\begin{equation}
\label{first}
\omega_1 \approx \pm \frac{c^2 k \rho}{c\rho \mp j \cos\theta } \approx
\pm c k + \frac{j}{\rho}\cos\theta
.
\end{equation}
This is equivalent to a sound wave propagating in the fluid moving with the velocity~$\jb/\rho$.

\subsection{Second Sound $\omega \ll ck$}
\label{incompress}
The binding energy of liquid helium is on the same order of magnitude as the quantum degeneracy energy (Fermi energy or BEC transition temperature). In $^4$He this corresponds to the approximate match between the vapor-liquid and the superfluid critical temperatures. The superfluidity energy scale in $^3$He (determined by the gap $\Delta$) is significantly smaller. This implies that the Landau critical velocity is much lower than the Fermi velocity as well as the speed of regular sound $v_L \ll v_F \sim c$. The low frequency flow is effectively incompressible (\cf\  \cite{LL6}) and the equations are much simplified. This is the case of the slow mode, when the density variations can be ignored:
\begin{equation}
\label{eq-incompress}
\begin{aligned}
\omega \frac{(\jb\jnbt)}{j} &=
& D k &\frac{(\jb\jnbt)}{j}
&+ k\left(w-\frac{C j}{\rho}\right)&\frac{(\kb\jnbt)}{k},
\\
\omega \frac{(\kb \jnbt)}{k} &=
& F k &\frac{(\jb\jnbt)}{j}
&+ 2k\left(w - \frac{j}{\rho}\right)\cos\theta &\frac{(\kb \jnbt)}{k}
.
\end{aligned}
\end{equation}
Let the $x$-axis run along \jb, and the wave-vector \kb\ lie in $xy$ plane, and denote the respective \jnbt\ projections as $\jntx,\jnty$. If we neglect $j\ll \rho w$, then the Eqs.\eqref{eq-incompress} take the form
\begin{equation*}
\begin{aligned}
\omega \jntx &=
& \left(\frac{4u}{3} + w\right)k\cos\theta &\jntx
&+ wk\sin\theta&\jnty, \\
\omega \jnty &=
& \frac{2u}{3}k\sin\theta&\jntx
&+ w k\cos\theta &\jnty.
\end{aligned}
\end{equation*}
This system has the following eigenvalues\footnote{The same eigenvalues can also be obtained for the full system \eqref{Matrix} in the principal order after cumbersome transformations.} and eigenvectors:
\begin{equation}
\label{second}
\begin{aligned}
\omega_2 &=
wk \cos\theta
+ \frac{2 u k }{3} \cos\theta 
\pm \frac{2k}{3} \sqrt{u^{2} \cos^2\theta + \frac{3}{2} u w \sin^2\theta},
\\
\begin{pmatrix}
\jntx\\[.2em]
\jnty
\end{pmatrix}
&\propto
\begin{pmatrix}
3 w \sin\theta\\[.2em]
-2 u \cos\theta \pm \sqrt{4 u^{2} \cos^2\theta + 6 u w \sin^2\theta}
\end{pmatrix}.
\end{aligned}
\end{equation}
As usual, two solutions and the $\pm$ sign correspond to the time reversal symmetry of the underlying equations.

\section{Discussion}
We have obtained exact linear-dispersion relations \eqref{first},\eqref{second} for two acoustic modes in overcritical Fermi superfluid. It is interesting, that no material constants (except the regular sound velocity $c$) enter these universal equations.

The usual (first) sound is only slightly affected \eqref{first} by small overcritical flow. The slow mode owes its very existence to the normal component and has distinctive angular behavior which reflects strongly anisotropic microscopic quasiparticle distribution. It is natural to analyze the second sound properties in the frame of reference of the reservoir walls (or equivalently, of the normal component). This corresponds to the Galilean transformation from \eqref{second} to
\begin{equation}
\label{secondn}
\omega_{2}-(\kb\wb)=
\frac{2 u k }{3} \cos\theta 
\pm \frac{2k}{3} \sqrt{u^{2} \cos^2\theta + \frac{3}{2} u v_L \sin^2\theta}.
\end{equation}
\begin{wrapfigure}{r}{0pt}
\begin{overpic}[width=0.3\linewidth]{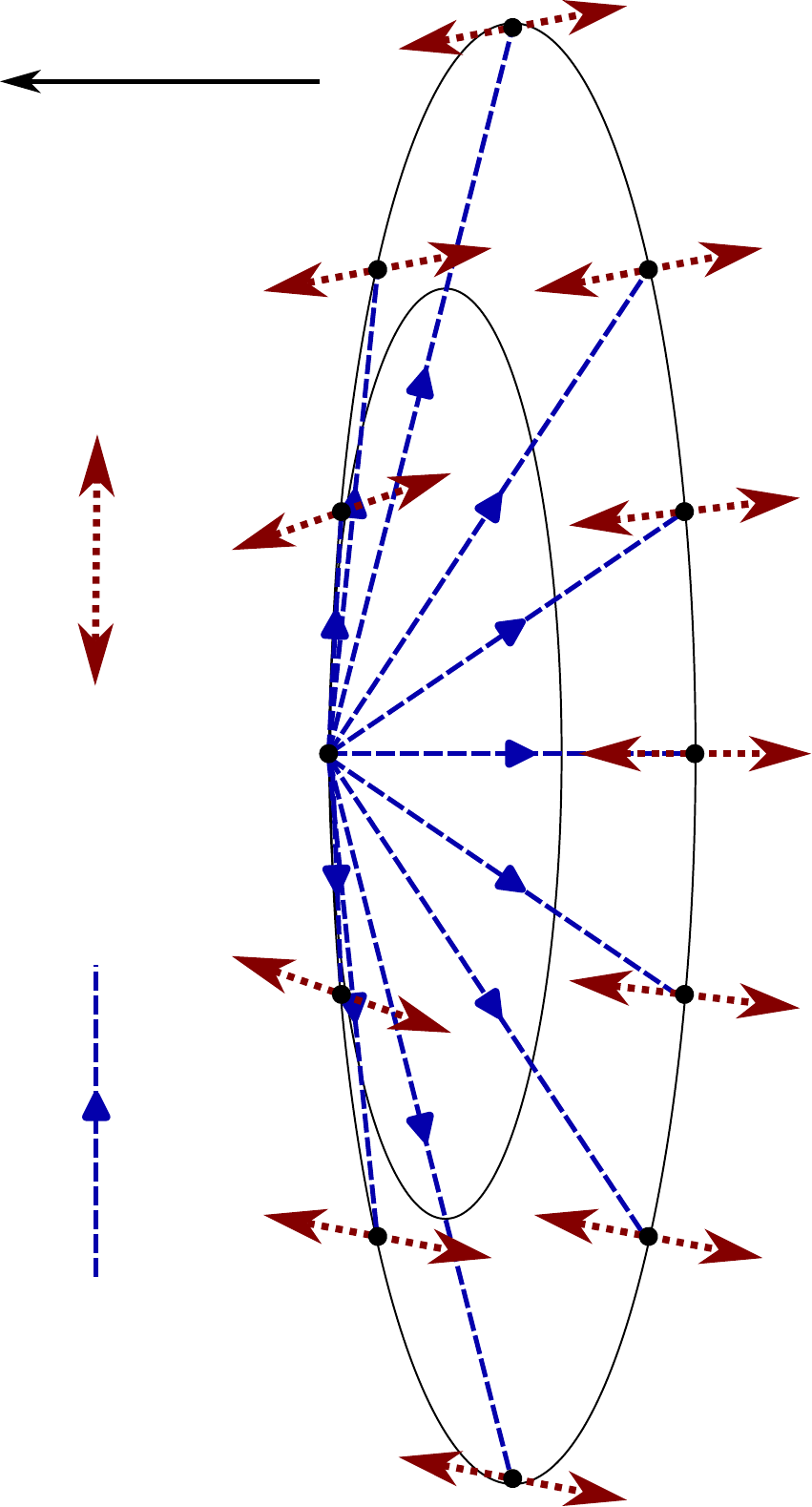}
\put(12,97){\vsb}
\put(17,48){$O$}
\put(10,62){\jnbt}
\put(10,25){ray}
\end{overpic}
\caption{Propagation of the second sound against the superfluid wind \vsb\ relative to the normal component from the origin $O$: elliptical (ellipsoidal in real three dimensions) wave-fronts, \jnbt\ polarization directions, and the straight rays.}
\label{propagation}
\end{wrapfigure}
A superfluid ``wind'' \vsb\ is present in this frame of reference, the second sound propagates {\em against} the wind: oblate ellipsoidal wave-fronts are confined to the right half-space in Fig.\ref{propagation}. Polarization of the second sound (relative alignment of \jnbt\ and \kb\ vector directions) is transversal for the wave-vector parallel to the wind (this is actually a singular point with zero velocity), and purely longitudinal in the opposite direction. For the wave vectors not in the immediate vicinity of these two distinct directions, the $uv_L$ term in the square root in \eqref{secondn} is dominant. The sound velocity is then approximately equal to $\sqrt{2 uv_L /3}$ and perpendicular to \vsb, while \jnbt\ is almost collinear with \vsb. The wave is therefore also transversal for $\theta\sim\pi/2+\sqrt{2u/3v_L}$ and almost longitudinal for $\pi-\theta \gtrsim u/v_L$.

The concept of second sound in superfluid $^4$He as a temperature wave without significant density variations is a consequence of anomalous smallness of the thermal expansion coefficient. At low temperature the normal component in overcritical Fermi superfluid carries no entropy and the second sound considered here is not a temperature wave. It is also not a density wave, but for a different reason: the first sound is much faster and the fluid compressibility can be neglected.

Obtained dispersion relations are valid for sufficiently long waves. The wavelength should be compared with the quasiparticle mean free path $l$ which grows at low temperatures \cite{chuck}. Similar arguments can be used to  estimate its behavior for Bogolyubov-Fermi quasiparticles
\begin{equation}
l
\sim \frac{p_F}{j_0 \sigma_0}\left(\frac{T_0}{T}\right)^2
\sim \frac{ \hbar \Delta u^{1/2} p_F^{1/2} }{T^2 m^{1/2}}
,
\end{equation}
where $j_0/p_F$ is the total quasiparticle count and $\sigma_0\sim \hbar^2/p_F^2$ is the bare scattering cross-section. The hydrodynamics equations are applicable for $kl\ll 1$. Damping will be high for short waves $kl\sim 1$. Possibility of the yet shorter $kl\gg 1$ wave propagation (similar to 0-sound in Fermi liquids) needs further investigation.

Present analysis deals with homogeneous background overcritical flow. It would be interesting to look for similar phenomena in nonuniform case, particularly in the vicinity of a vortex line and within the vortex lattice, or investigate the flow past an obstacle.

\section{Acknowledgements}
Fruitful discussions with V.I.Marchenko, S.Refaeli-Abramson, E.V.Surovtsev are gratefully appreciated. This work was partially supported by the MOIA grant \#140459.

\end{document}